\RequirePackage[displaymath]{lineno}
\documentclass[aps,prc,twocolumn,superscriptaddress,floatfix,showpacs,nofootinbib]{revtex4-1}
\usepackage{dcolumn}% Align table columns on decimal point
\usepackage{bm}% bold math
\usepackage{float}
\usepackage[]{graphicx}  % remove 'demo' option for your real document
\usepackage{booktabs}
\usepackage{tabularx}
\usepackage{amsmath}
\usepackage{xcolor}
\usepackage{multirow}
\usepackage[colorlinks,citecolor=blue,urlcolor=blue,linkcolor=blue]{hyperref}

\newcommand{\snn}{\sqrt{s_{\rm NN}}}
\newcommand{\dr}{$\Delta r_{\rm np}$}
\newcommand{\Pb}{$^{208}$Pb}

\newcommand{\Ca}{$^{48}$Ca}
\newcommand{\Au}{$^{197}$Au}

\newcommand{\PbPb}{$^{208}$Pb+$^{208}$Pb}
\newcommand{\AuAu}{$^{197}$Au+$^{197}$Au}
\newcommand{\UU}{$^{238}$U+$^{238}$U}
\newcommand {\RuRu}	{$^{96}_{44}$Ru+$^{96}_{44}$Ru}
\newcommand {\ZrZr}	{$^{96}_{40}$Zr+$^{96}_{40}$Zr}
\newcommand {\Nch}      {N_{\rm ch}}
\newcommand {\mean}[1]  {\langle #1\rangle}

\newcommand {\rnp}  {\Delta r_{\rm np}}

\newcommand{\trento}{T\raisebox{-0.5ex}{R}ENTo}

\begin{document}

\title{Determining the neutron skin thickness by relativistic semi-isobaric collisions}

\author{Qi Liu}
\affiliation{School of Physics, Peking University, Beijing 100871, China}

\author{Shujun Zhao}
\affiliation{School of Physics, Peking University, Beijing 100871, China}

\author{Hao-jie Xu\footnote{%
        Corresponding author: haojiexu@zjhu.edu.cn}}
\affiliation{School of Science, Huzhou University, Huzhou, Zhejiang 313000, China}
\affiliation{Strong-Coupling Physics International Research Laboratory (SPiRL), Huzhou University, Huzhou, Zhejiang 313000, China.}
\affiliation{Shanghai Research Center for Theoretical Nuclear Physics, NSFC and Fudan University, Shanghai 200438, China.}

\author{Huichao Song\footnote{%
        Corresponding author: huichaosong@pku.edu.cn}}
\affiliation{School of Physics, Peking University, Beijing 100871, China}
\affiliation{Center for High Energy Physics, Peking University, Beijing 100871, China}
\affiliation{Collaborative Innovation Center of Quantum Matter, Beijing 100871, China}

\begin{abstract}
The neutron skin thickness of the benchmark nucleus \Pb\ is crucial for our understanding of the equation of state of nuclear matter. In this paper, we discuss the effect of the neutron skin on the flow ratio observables in the semi-isobaric collisions \PbPb\ and \AuAu\ using iEBE-VISHNU  hydrodynamic simulations.
Our results suggest that \Pb\ and \Au\ should have the same magnitude of neutron skin thickness to describe the anisotropic flow ratios between the semi-isobaric systems.
Our method provides an unconventional way to determine the neutron skin with the existing relativistic heavy ion collision data.
\end{abstract}

\maketitle
%\pacs{05.70.Jk, 25.75.Gz, 25.75.-q, 25.75.Nq}

\section{Introduction}\label{Introduction}

The equation of state (EoS) of nuclear matter governs the general properties of the nuclear force over several orders of magnitude, from nuclei to neutron stars~\cite{Brown:2000pd,Horowitz:2000xj,Li:2008gp,Chen:2010qx}.
The symmetry energy in the EoS encodes the energy associated with the neutron-proton asymmetry.
Measurements of the neutron skin thickness \dr\ can provide valuable information about the nuclear symmetry energy~\cite{Bartel:1982ed,Machleidt:1989tm,AlexBrown:1998zz,Furnstahl:2001un,Chen:2005ti,RocaMaza:2011pm,Tsang:2012se,Horowitz:2014bja}.
The neutron skin of the benchmark nuclei \Pb\ (\dr=$0.28\pm0.07$ fm) and \Ca\ (\dr=$0.121\pm0.026\pm0.024$ fm) were recently been measured by the lead radius experiment (PREX-II) and the calcium radius experiment (CREX) with parity violating electron nucleus scattering processes~\cite{Adhikari:2021phr,CREX:2022kgg}.
Based on the energy density functional calculations, the PREX-II data lead to a stiff EoS, while the CREX data favor a softer EoS~\cite{Reed:2021nqk}.
These results challenge our current understanding of the EoS of nuclear matter, which lead to extensive studies in low energy nuclear physics in the past two years~\cite{Yuksel:2022umn,Zhang:2022bni}.
In general, the measurement of the mass radii can be used directly to determine the neutron skin thickness of such nuclei, given that their charge radii can be accurately measured.
The photoproduction of neutral $\rho$ mesons has been found to be sensitive to the mass radii of the colliding nuclei, and this method has been applied in relativistic heavy ion collisions~\cite{Alvensleben:1970uw,STAR:2022wfe}.
The neutron skin of \Au\ (\dr=$0.17\pm0.03\pm0.08$ fm) has been measured by the STAR Collaborations with the $\rho^0\rightarrow\pi^+\pi^-$ photoproduction method in ultra-peripheral Au+Au collisions (UPC) at $\snn=200$ GeV~\cite{STAR:2022wfe}.

Recently, more unconventional methods have been proposed for neutron skin thickness measurements in relativistic heavy ion collisions~\cite{Li:2019kkh,Zhang:2021kxj,Nijs:2021kvn,Jia:2022qgl,
Zhang:2022fou,Nie:2022gbg,Xi:2023isk,Cheng:2023ucp,vanderSchee:2023uii,Giacalone:2023cet}.
In the heavy ion collisions at top RHIC and the LHC energies,  the quark gluon plasma (QGP), a decoupled phase of quarks and gluons, has been created~\cite{Adams:2005dq,Adcox:2004mh,ALICE:2010suc,Gyulassy:2003mc,Muller:2012zq},
which can be successfully described by relativistic hydrodynamics simulations with a small specific shear viscosity~\cite{Song:2010mg,Gale:2013da,Song:2013gia,Heinz:2013th,Song:2017wtw,JETSCAPE:2020mzn,JETSCAPE:2020shq}.
It was also found that the final state observables, such as multiplicity, mean transverse momentum and anisotropic flow, are intrinsically related to the initial state and thus sensitive to the structure of the colliding nuclei~\cite{Ollitrault:1992bk,Heinz:2013th,Song:2017wtw,Bally:2022vgo}.
These observables are subject to less uncertainty from the strong interactions in quantum chromodynamic theory than the observables in low-energy hadronic collisions.
However, it is difficult to directly distinguish the subtle nuclear structure effects in the single collision system due to the uncertainties in the bulk properties for the QGP evolution, although some efforts have been made recently~\cite{Giacalone:2023cet,Bally:2022vgo}.
The RHIC isobar runs with \ZrZr\ and \RuRu\ collisions at $\snn=200$ GeV,
originally designed to search for the Chiral Magnetic Effect (CME),
provide unique opportunities to probe the nuclear structures of the colliding nuclei from the early stages~\cite{Xu:2017zcn,STAR:2021mii,Li:2019kkh,Xu:2021qjw,Zhang:2021kxj,Nijs:2021kvn,Jia:2022qgl,Zhang:2022fou,Nie:2022gbg,Xi:2023isk,Cheng:2023ucp,vanderSchee:2023uii}.
This is because the uncertainties from the bulk properties of the QGP can be significantly reduced by the observable ratios between the two collision systems~\cite{STAR:2021mii,Xu:2021uar}.
Due to the large statistics of isobar collisions, the differences between the two collision systems can be measured with high precision.
Previous studies have proposed that the differences in multiplicity and mean transverse momentum between the isobar collisions can be used to probe the neutron skin and the symmetry energy slope parameter~\cite{Li:2019kkh,Xu:2021uar}. These studies also suggested that the elliptic flow measurements can also determine the proper nuclear structures of the isobaric nuclei~\cite{Xu:2021vpn,Zhang:2021kxj}.

\begin{figure}[hbt!]
        \includegraphics[width=0.44\textwidth]{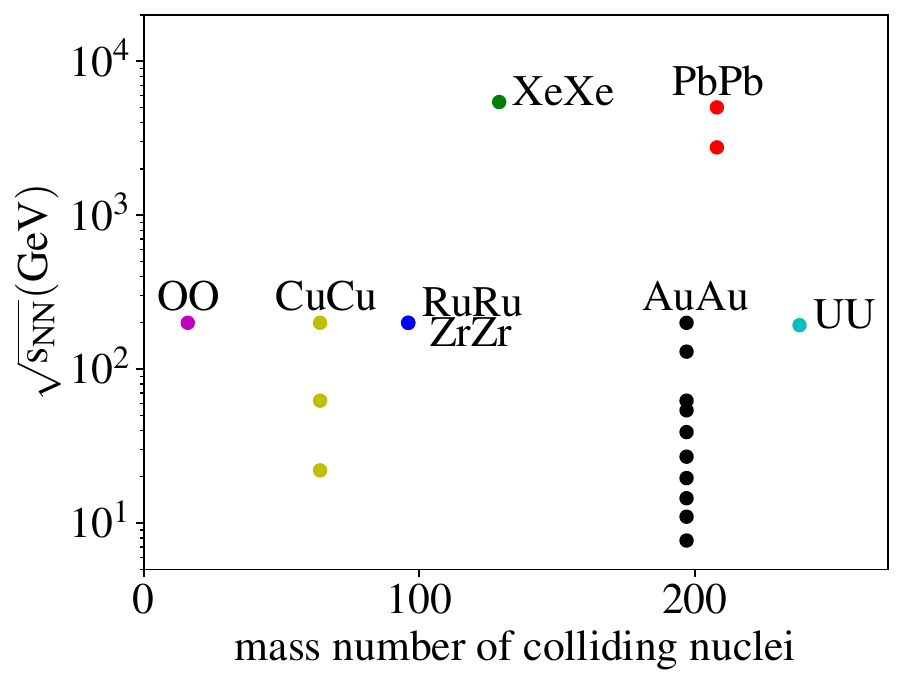}
	\caption{(Color online) Different symmetric collision systems at RHIC and the LHC.~\label{fig:collsys}}
\end{figure}

Because of the advantages of isobar collisions, one might expect to use such a method to determine the neutron skin of the benchmark nuclei \Pb \ with the rich soft data from Pb+Pb collisions at the LHC~\cite{ALICE:2010suc,ALICE:2012eyl}. Currently, no suitable isobar partner for \Pb \ has been collided in relativistic heavy ion collisions. Figure~\ref{fig:collsys} summarizes the symmetric collisions systems
at RHIC and the  LHC .
The only existing data for isobar collisions in relativistic heavy ion collisions are the \RuRu\ and \ZrZr\ collisions at the top RHIC energy~\cite{STAR:2021mii}. If we relax the restriction and consider colliding nuclei with similar baryon numbers, called semi-isobaric, we can discuss the ratio observables in such semi-isobaric collisions.
For example, the ratio observables between \AuAu\ and \UU\ collisions have been used to discuss the effect of nuclear deformation~\cite{Jia:2021wbq}.
For the purpose of determining the neutron skin thickness of \Pb, the \Au\ is a good choice with a large amount of collision experiment data and small deformation effect~\cite{STAR:2008med,STAR:2004jwm,STAR:2013qio,Moller:1993ed}.
In this paper, we calculate the anisotropic flow $v_2$ and $v_3$ of \AuAu\ at $\snn=200$ GeV and \PbPb\ at $\snn=2.76$ TeV with different neutron skin sizes and compare them with the measured data of the flow observables. We aim to explore the effect of the neutron skin on the anisotropic flow differences between the semi-isobaric systems.

The paper is organized as follows. Section~\ref{sec:model} gives a brief description of the models and the nuclear density profiles used in this work. Section~\ref{sec:results} discusses the effects of neutron skin on the flow observables in the semi-isobar collisions, \PbPb\ collisions and \AuAu\ collisions.  Sec.~\ref{sec:summary} summarizes and concludes this paper.

\section{Model setup}\label{sec:model}

The effect of neutron skin on the final observables in relativistic heavy ion collisions is studied using an iEBE-VISHNU model~\cite{Song:2010aq,Shen:2014vra}.
iEBE-VISHNU~\cite{Shen:2014vra,Song:2010aq} is a state-of-the-art hybrid model with (2+1)-dimensional viscous hydrodynamics~\cite{Song:2007ux,Song:2007fn} to describe the expansion of QGP matter, followed by a hadron cascade model (UrQMD) to simulate the evolution of the subsequent hadronic matter~\cite{Bass:1998ca,Bleicher:1999xi}. For more details on the model and parameter set-ups, please refer to~\cite{Shen:2014vra,Xu:2016hmp,Zhao:2017yhj}.
Here, we implement \trento\ initial condition model~\cite{Moreland:2014oya} to start the hydrodynamic simulations.
The nuclear density in \trento\ is described by a Woods-Saxon (WS) distribution,
\begin{eqnarray}
	\rho(r) = \frac{\rho_0}{1+\exp{(\frac{r-R}{a_{0}})}},
\label{eq:one}
\end{eqnarray}
with
\begin{eqnarray}
R = R_0(1+\beta_2 Y^{0}_{2}(\theta,\phi)+ \beta_3 Y^{0}_{3}(\theta,\phi) + ...),
\label{eq:two}
\end{eqnarray}
where \(\rho_0\) is the nuclear saturation density, \(a\) is the diffuseness parameter, \(R_0\) is the radius parameter, and \(\beta_2\) and \(\beta_3\) are the nuclear deformation parameters. In this study, we set \(\beta_n\)  of \Au\ and \Pb\ to zero because the nuclear deformation parameters are expected to be small for both nuclei.

The nuclear densities can in principle be calculated with some fundamental nuclear structure theories such as energy density functional theory, and the neutron skin is strongly correlated with the symmetry energy slope parameter. For simplicity, we follow the Woods-Saxon framework.
The low-energy nuclear experiments indicate that the neutron skin for most of the nuclei is halo-type~\cite{Trzcinska:2001sy}. The halo-type neutron skin thicknesses of \Pb\ and \Au\ are constructed by using different diffuseness parameters $a$ for the proton and mass densities, keeping the radius parameter $R_{0}$ fixed. The WS parameters for the nuclear mass densities are listed in Tab.~\ref{isobar_ws_table}, the corresponding WS parameters of the proton density are the same as those of the mass density in the case of \dr=$0$ fm.
In this study, we used three sets of neutron skin thicknesses for both Pb and Au, i.e. \dr$=0$ fm, \dr$=0.17$ fm, and \dr$=0.28$ fm, respectively.
The \dr$=0.28$ fm are taken from the PREX-II experiment for the \Pb\ nuclei by electroweak parity-violating scattering processes~\cite{Adhikari:2021phr}.
The \dr=$0.17$ fm are taken from the STAR measurement for the \Au\ nuclei by relativistic Au+Au collisions at $\snn=200$ GeV~\cite{STAR:2022wfe}.

\begin{figure*}[hbt!]
        \includegraphics[width=0.44\textwidth]{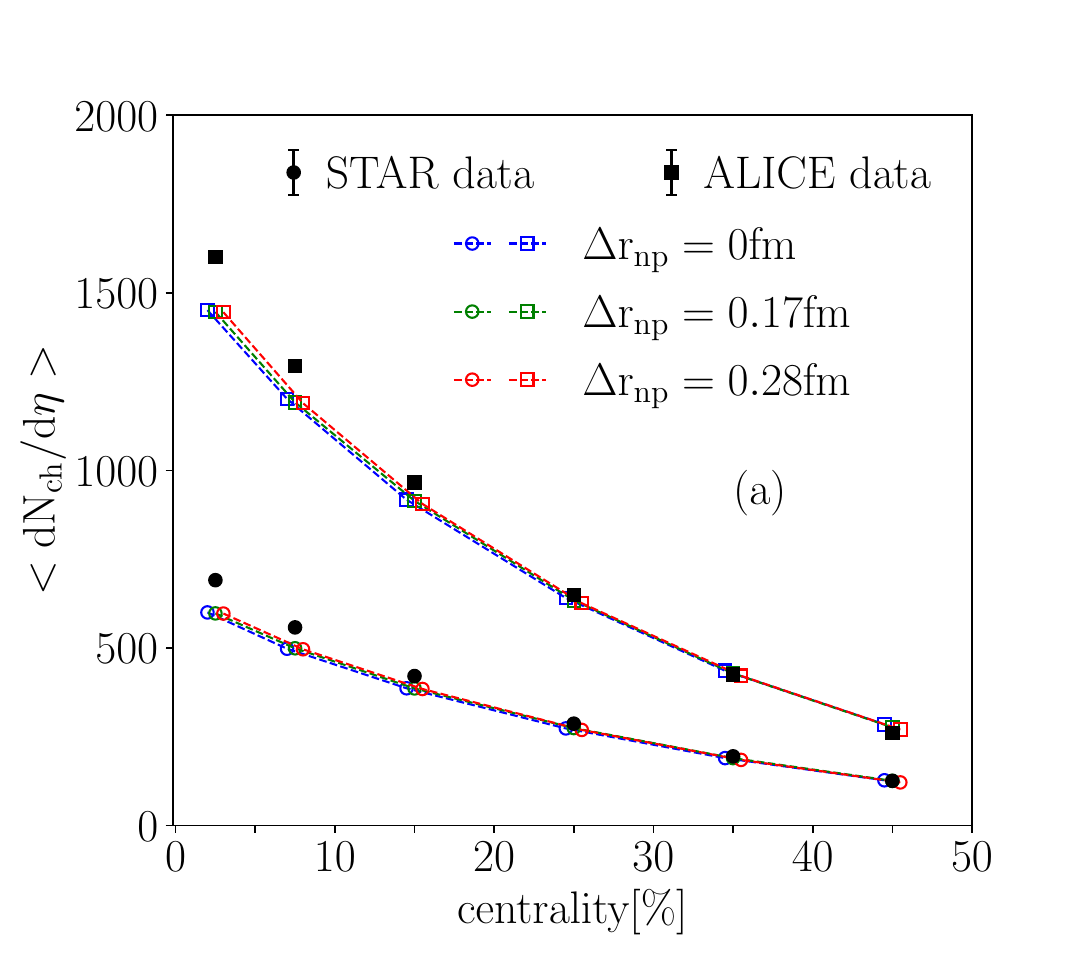}
        \includegraphics[width=0.44\textwidth]{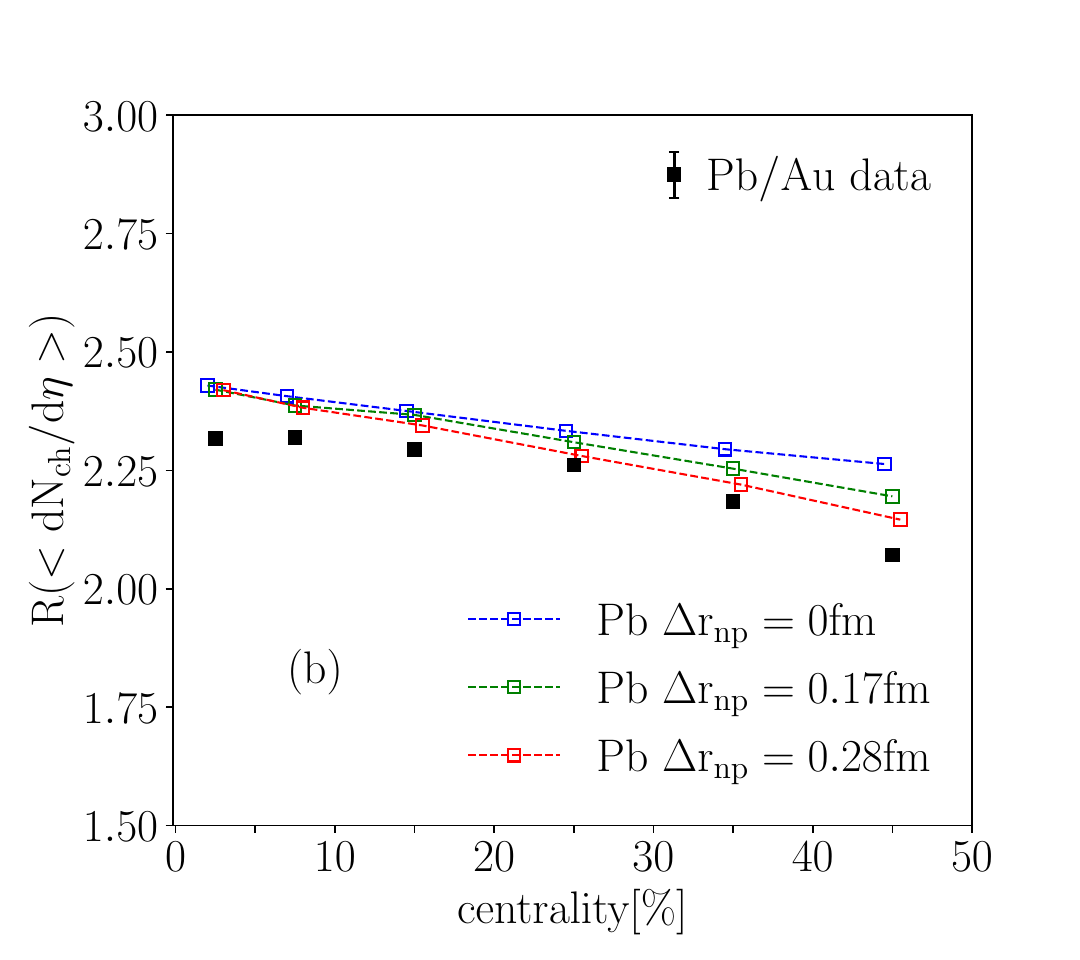}
	\caption{(Color online) Charged particle production with different sizes of neutron skin in (a) \PbPb\ collisions at $\snn=2.76$ TeV,
    \AuAu\ collisions at $\snn=200$ GeV, and (b) their ratios.~\label{fig:dNdeta}}
\end{figure*}

\begin{table}[hbt!]
	\caption{The WS parameter for \Pb\ and \Au\ with different values of the halo-type neutron skin thickness.~\label{isobar_ws_table}}
	\centering{}
  \begin{tabular}%{|c|c|c|c|c|}
	  {p{1.6cm}p{1.4cm}p{1.4cm}p{1.4cm}p{1.4cm}}
	\hline
	\dr  & \multicolumn{2}{c}{\Pb} & \multicolumn{2}{c}{\Au}  \\
      (fm) &  $R_0$(fm) &  $a$(fm) &  $R_0$(fm) & $a$(fm)\\
    \hline
    0  & 6.62 & 0.546 & 6.38 & 0.535\\
    \hline
    0.17 &  6.62 & 0.6177 & 6.38 & 0.605\\
    \hline
    0.28 & 6.62 & 0.661 & 6.38 & 0.647\\
    \hline
  \end{tabular}
\end{table}

\begin{figure*}[hbt!]
        \includegraphics[width=0.44\textwidth]{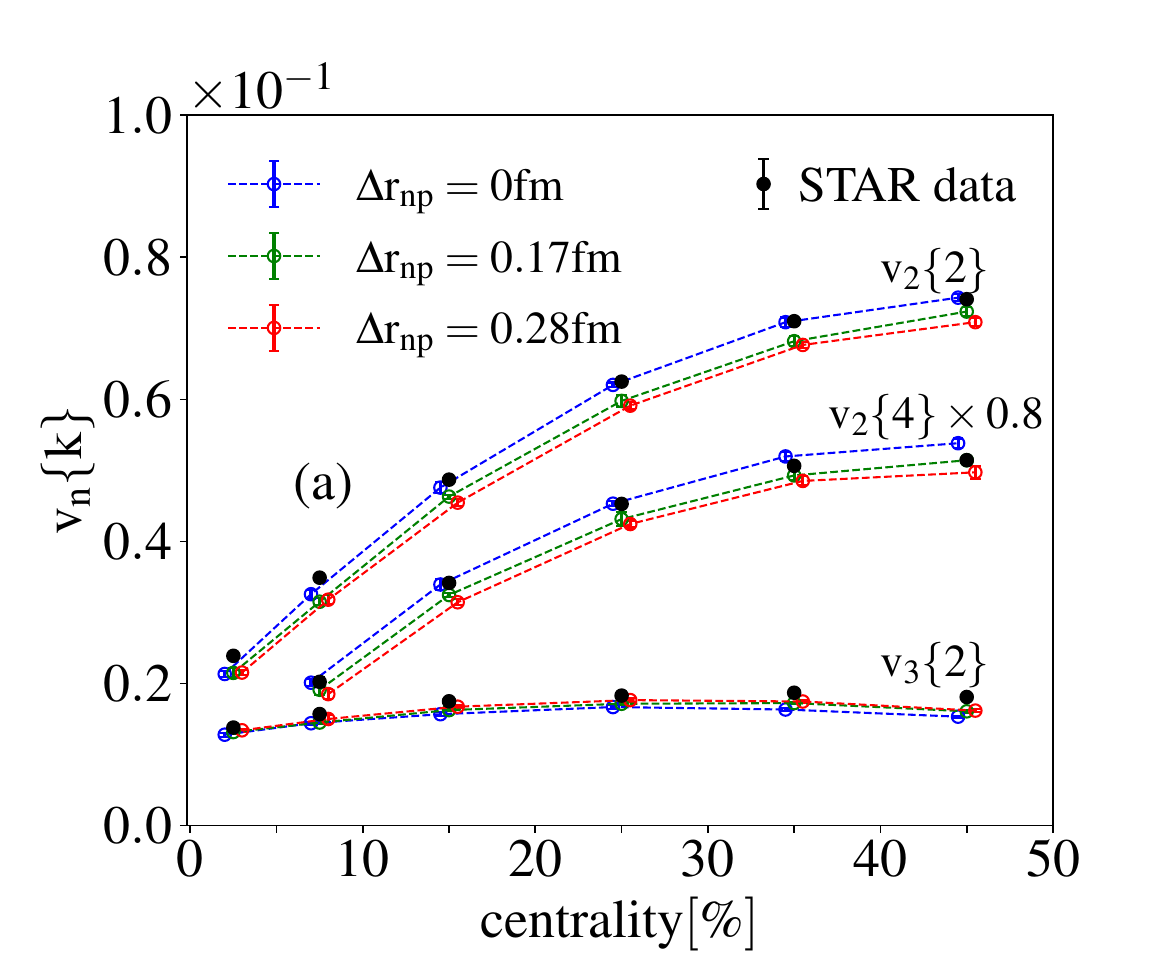}
        \includegraphics[width=0.44\textwidth]{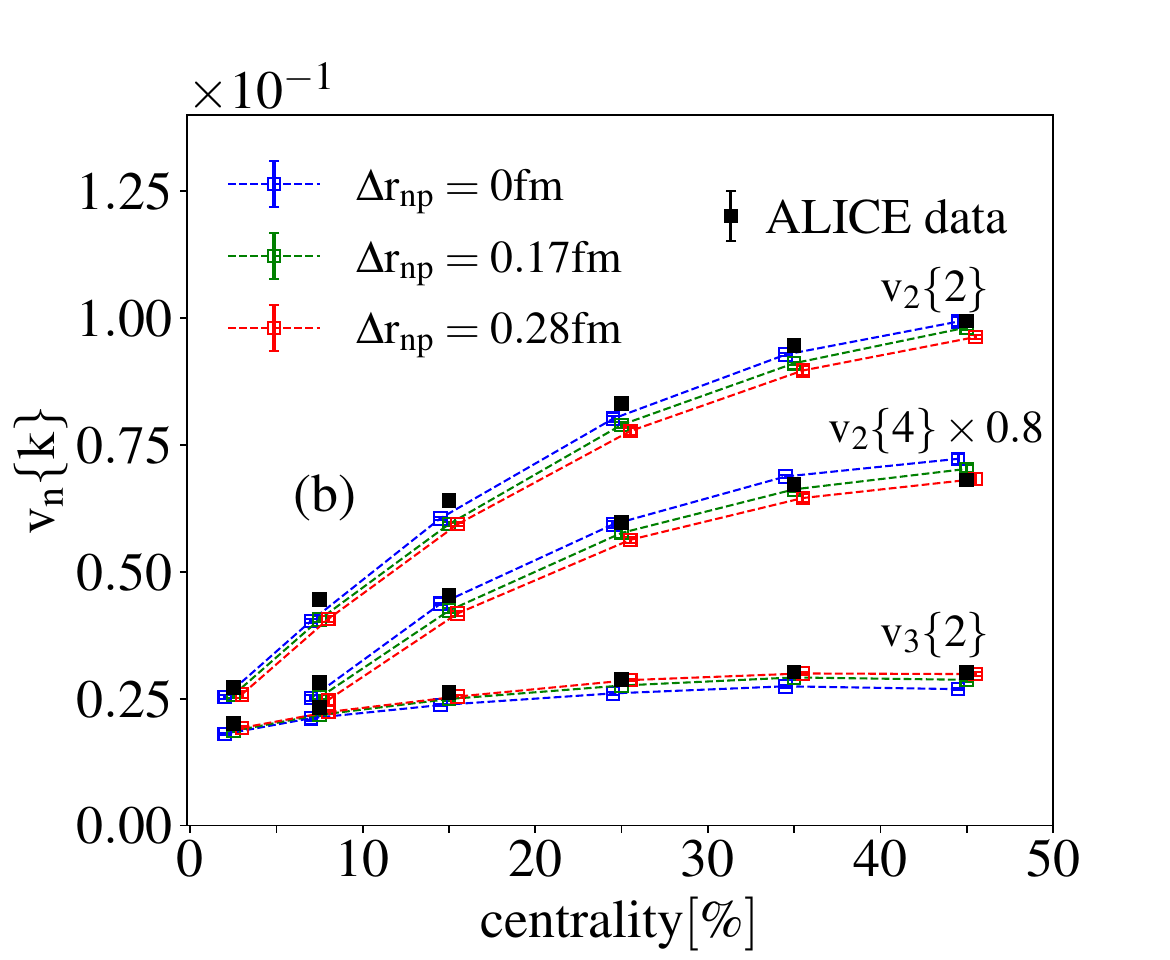}
	\caption{(Color online)  Flow harmonics $v_{2}\{2\}$, $v_{2}\{4\}$, and  $v_{3}\{2\}$ in Pb+Pb collisions at $\snn=2.76$ TeV and Au+Au collisions at $\snn=200$ GeV, calculated from iEBE-VISHNU hydrodynamic simulations with different sizes of neutron skin for the colliding nuclei.~\label{fig:individualflow}}
\end{figure*}

For the following iEBE-VISHNU simulations for \PbPb\ collisions at $\snn=2.76$ TeV  and \AuAu\ collisions at $\snn=200$ GeV, most of the parameters are taken from previous Bayesian analysis~\cite{Bernhard:2019bmu}, which are extracted from the spectra and flow data  measured in \PbPb\ collisions at $\snn=2.76$ TeV. 
We use the improved \trento\ version with nucleon constituent scenario, all the model parameters are listed in Tab.~\ref{tab:para}.  
Here, we have only changed the normalization factor Norm=$5.35$ GeV and the inelastic cross section $\sigma_{\rm NN}=4.23$ fm$^2$~\cite{PHENIX:2015tbb} for a better description of the soft hadron data in \AuAu\ at  $\snn=200$ GeV at RHIC.
We also note that the hydrodynamic parameters are extracted with the \dr=$0$ fm in the simulations of Pb+Pb collisions, the values should be changed at some levels if the neutron skin effect is included.
Since we focus on the ratio observables described by the following Eq.(3) in this study, the main conclusions will not be largely influenced if other hydrodynamic parameters are used.

\begin{table}
 \caption{The parameters of iEBE-VISHNU simulaitons for {$^{208}$Pb+$^{208}$Pb}/{$^{197}$Au+$^{197}$Au} collisions.
 \label{tab:para}}
 \centering{}%
    \begin{tabular}%{ccccc} 
    {p{1.5cm}p{2.5cm}p{1.5cm}p{2.5cm}}
    \hline
         \multicolumn{2}{c}{{T\raisebox{-0.5ex}{R}ENTo}/freestream.} & \multicolumn{2}{c}{osu-hydro}  \\
    \hline
	    Norm  & $13.94$/$5.35$ GeV  & $(\eta/s)_{\rm min}$           & $0.081$   \\
	    $\sigma_{\rm NN}$  & $6.4$/$4.23$ ${\rm fm^2}$ & $(\eta/s)_{\rm slope}$           & $1.11$ GeV$^{-1}$    \\
     $k$  & $0.1978$  & $(\eta/s)_{\rm crv}$            & $-0.48$    \\
     $w$ & $0.956$ fm  & $(\eta/s)_{\rm hrg}$            & $0.5$    \\
	$p$     & $0.007$   &   	$(\zeta/s)_{\rm max}$         & $0.052$   \\
     $m$  & $6$    & $(\zeta/s)_{\rm width}$       & $0.022$ GeV   \\
     $v$   & $0.956$ fm     & $(\zeta/s)_{T_{\rm peak}}$    & $0.183$ GeV   \\
     $d_{\rm min}$  & $1.27$ fm       & $T_{\rm switch}$                & $0.151$ GeV   \\
     $\tau_{\rm fs}$ & $1.16$ fm/$c$ &                                             &    \\   
 \hline
 \end{tabular}
\end{table}

\begin{figure*}[hbpt]
        \includegraphics[width=0.44\textwidth]{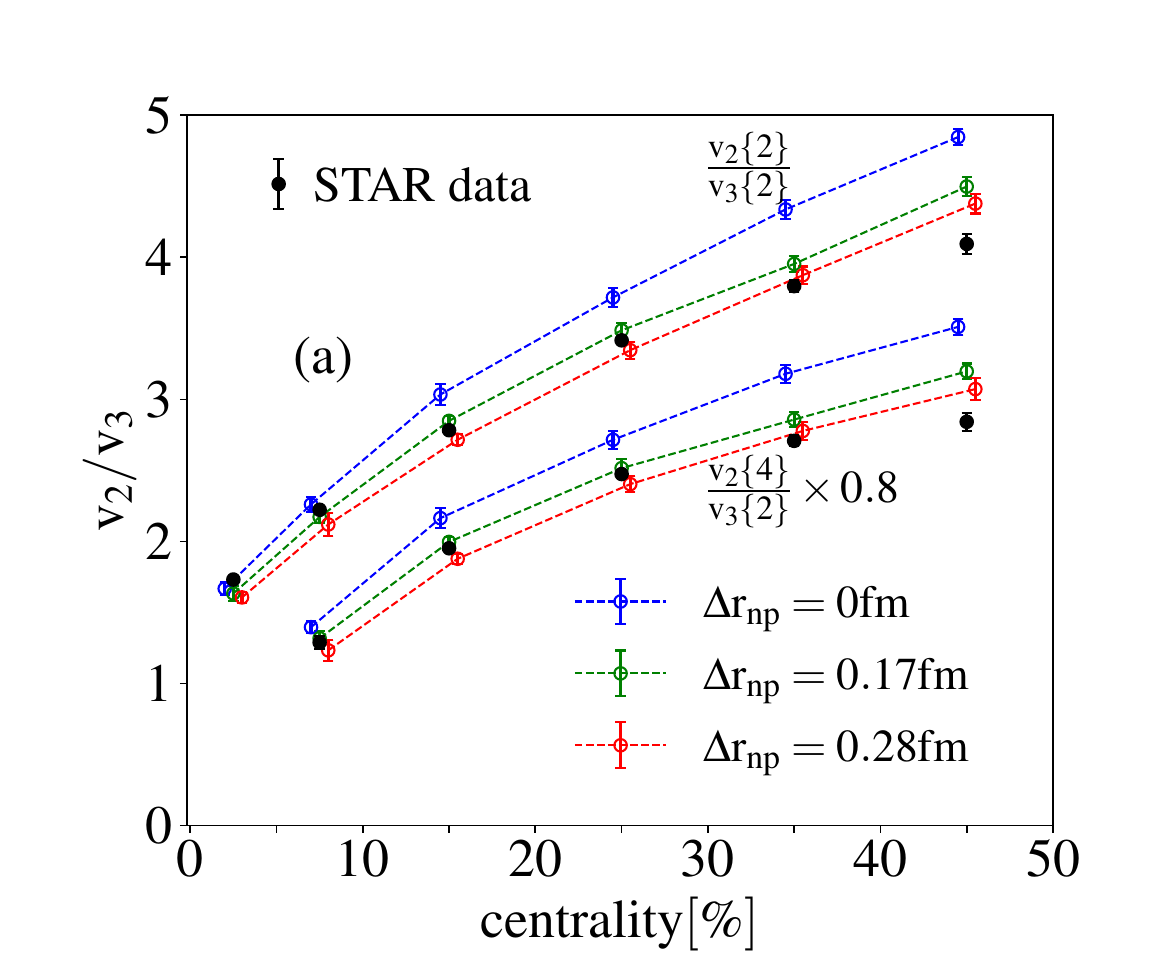}
        \includegraphics[width=0.44\textwidth]{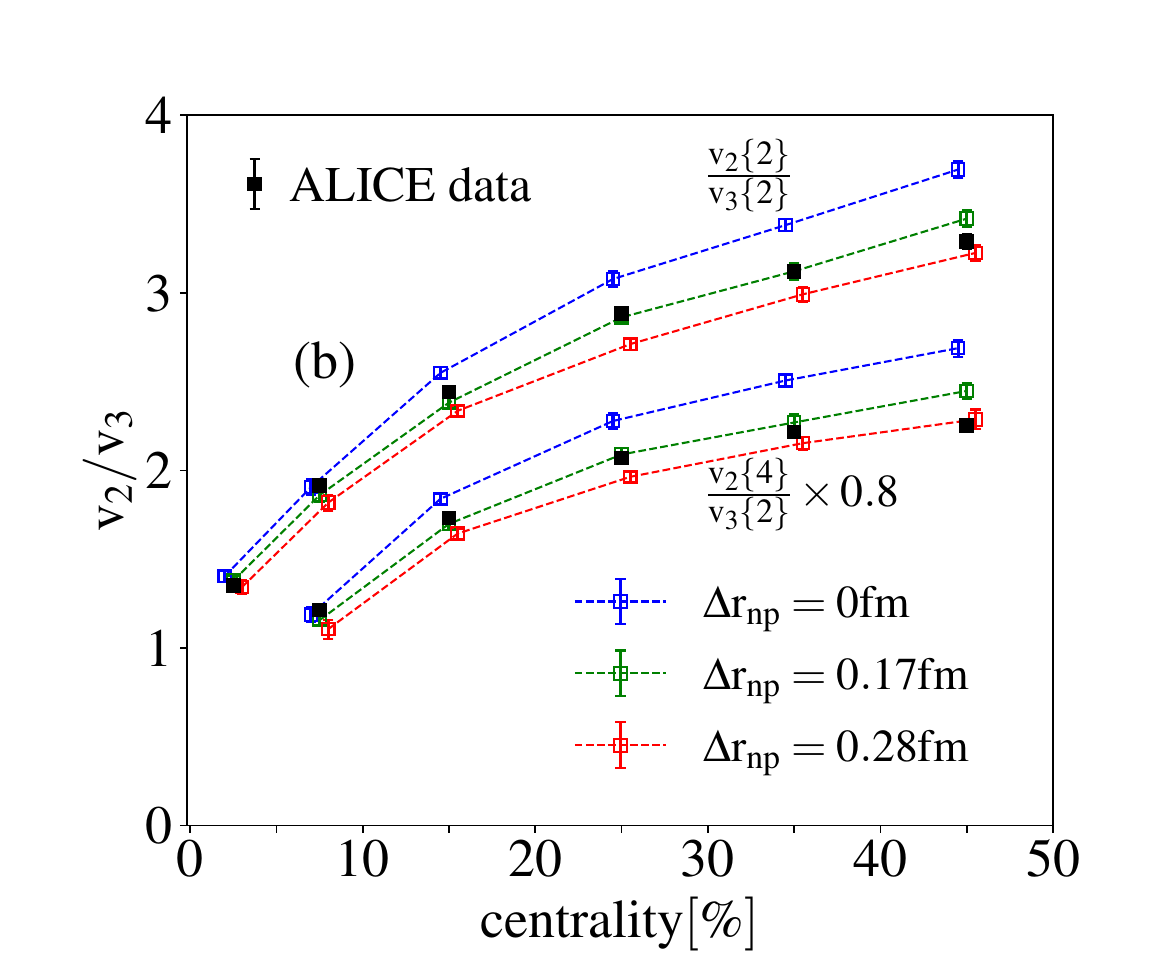}
	\caption{(Color online) The flow ratios of $v_{2}\{2\}/v_{3}\{2\}$ and $v_{2}\{4\}/v_{3}\{2\}$ in (a) Au+Au collisions at $\snn=200$ GeV and (b) Pb+Pb collisions at $\snn=2.76$ TeV from iEBE-VISHNU hydrodynamic simulations with different sizes of neutron skin for the colliding nuclei.~\label{fig:v2v3}}
\end{figure*}

In order to reduce the systematic uncertainties, we will focus on the ratios of the flow observables between the semi-isobaric systems (\PbPb\ at $\snn=2760$ GeV and \AuAu\ at $\snn=200$ GeV), which is defined as:
\begin{equation}
      R(X) \equiv \frac{X^{\rm PbPb}}{X^{\rm AuAu}},
\end{equation}
Here, the flow observables \(v_{2}\{2\}\), \(v_{2}\{4\}\) and \(v_{3}\{2\}\) are calculated by the standard Q-cumulant methods~\cite{Bilandzic:2010jr}.
The results are compared with the RHIC and LHC data~\cite{STAR:2008med,STAR:2004jwm,STAR:2013qio,ALICE:2016ccg}, and only the statistical uncertainties are considered for the data/model comparisons.

\section{Result and Discussion}\label{sec:results}

Figure~\ref{fig:dNdeta} (a) shows the mean multiplicity as a function of centrality for \AuAu\ collisions at $\snn=200$ GeV and \PbPb\ collisions at $\snn=2.76$ TeV, calculated from iEBE-VISHNU simulations with different sizes of neutron skin thickness. It demonstrates that the effect of the neutron skin on the mean multiplicity is considerably small, even for the ratios in the semi-isobaric collisions as shown in panel(b).
Previous studies in exactly isobar collisions with high statistics indicate that the mean multiplicities are sensitive to the neutron skin thicknesses and thus to the symmetry energy slope parameter~\cite{Li:2019kkh,Xu:2021vpn}.
However, the effect is only less than $1\%$ in most central isobar collisions, it is invisible in this study with current statistics, and have marginally differences at peripheral collisions.
We note that the uncertainties from the normalization factor cannot be eliminated for the $R(\mean{\Nch})$, since we are dealing with the collision system at different collision energies. Therefore, the multiplicity ratio is not a good observable to discuss the neutron skin effect in such semi-isobar collisions, and the mean transverse momentum $\langle p_{T}\rangle$ have the similar issue. Therefore, the exactly isobar collisions for \Pb\ and/or \Au\ are desired to demonstrate the effects. 

\begin{figure*}[hbt!]
        \includegraphics[width=0.44\textwidth]{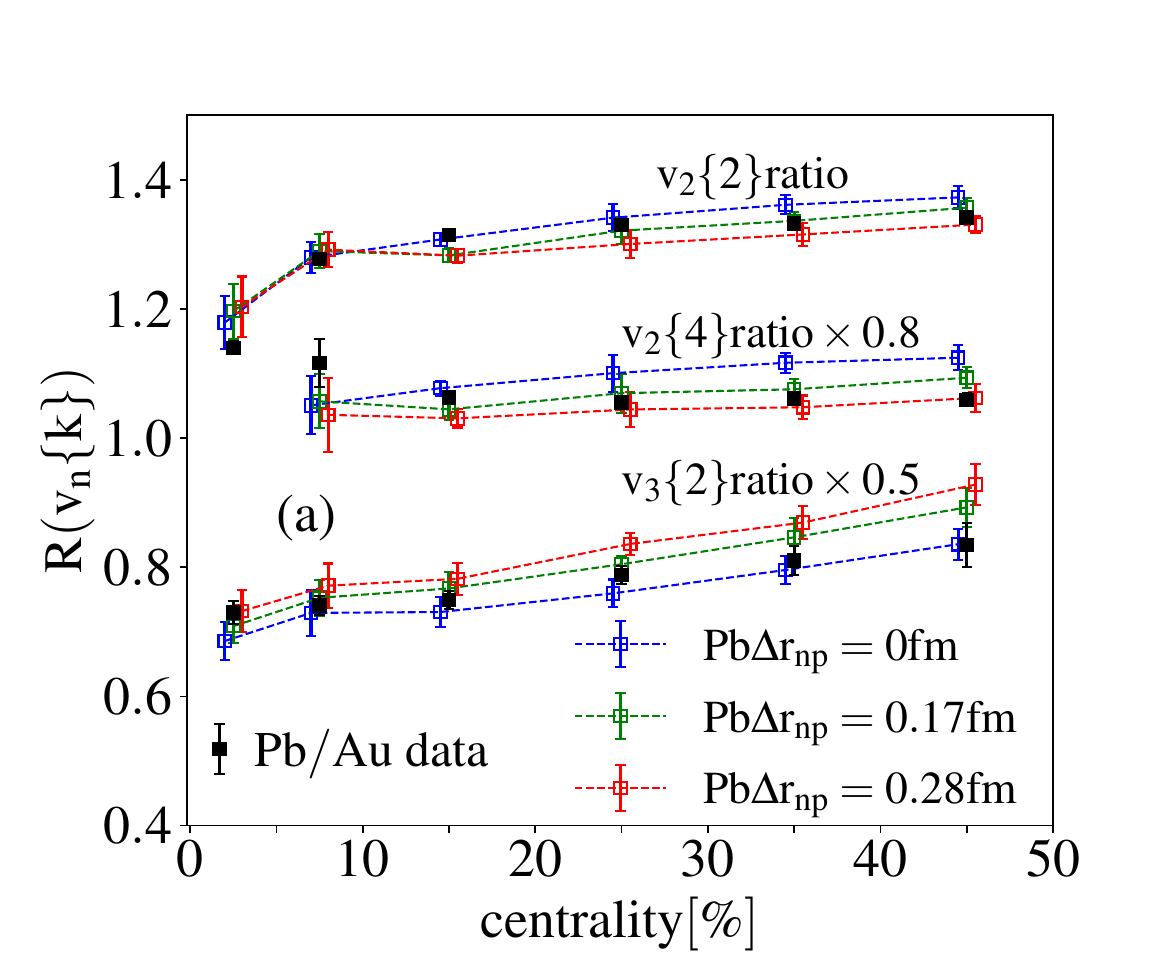}
        \includegraphics[width=0.44\textwidth]{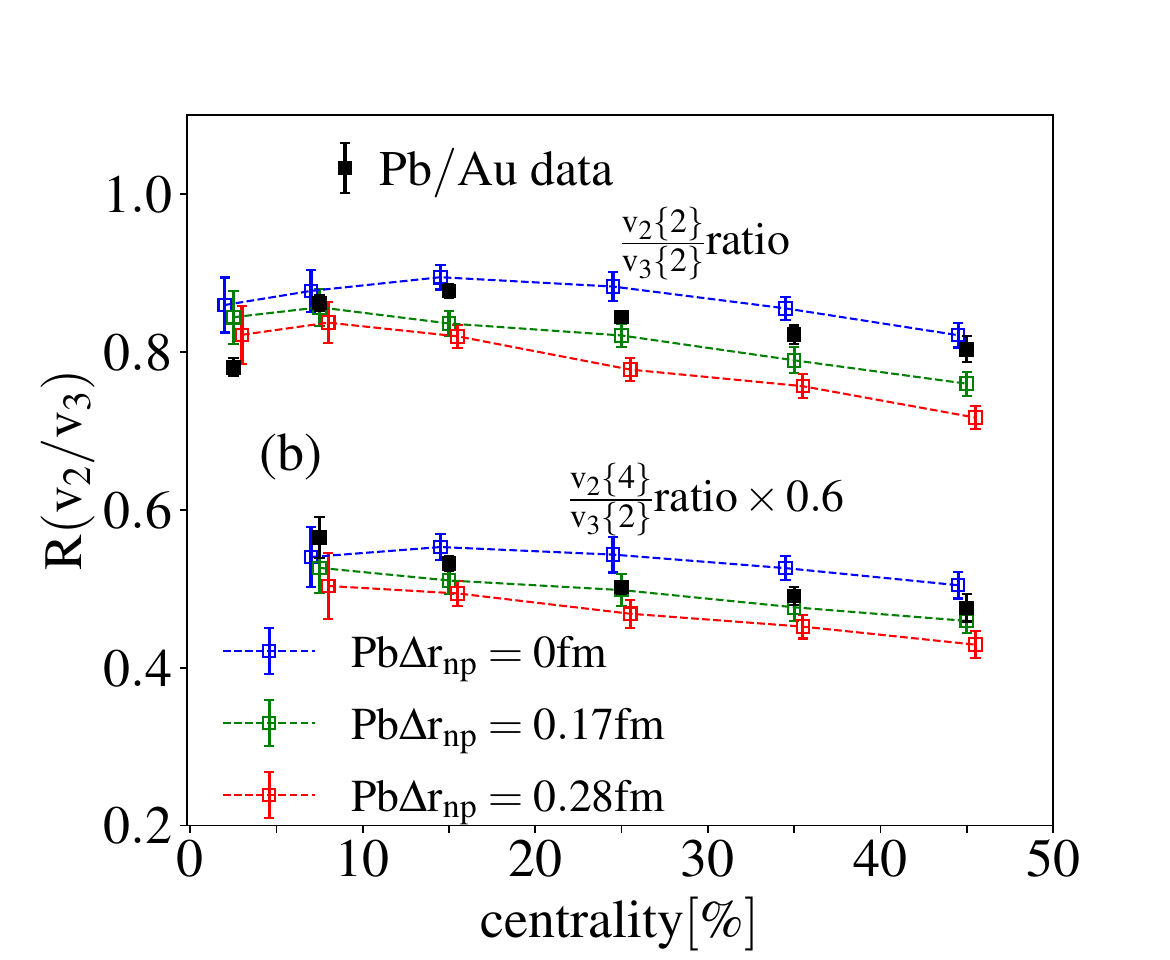}
	\caption{(Color online) The ratios of (a) individual flow observables $v_{2}\{2\}$, $v_{2}\{4\}$, $v_{3}\{2\}$  and (b) their double ratios in semi-isobaric collisions, calculated from iEBE-VISHNU hydrodynamic simulations with different sizes of neutron skin for the \Pb\ nuclei. The neutrons skin for \Au\ is fixed to $\rnp=0.17$ fm.~\label{fig:isobarratio}}
\end{figure*}

The halo-type neutron skin is expected to give opposite contributions to $v_2$ and $v_3$ in non-central relativistic heavy ion collisions~\cite{Xu:2021vpn,Zhang:2021kxj}. Figure~\ref{fig:individualflow} shows the effect of the neutron skin on $v_2$ and $v_3$ in semi-isobaric collisions.
Results from hydrodynamic simulations without neutron skin contributions work well for $v_{2}\{2\}$ and $v_{2}\{4\}$ in both Pb+Pb and Au+Au collisions, but are underestimated for $v_{3}\{2\}$.
These results are consistent with the previous Bayesian simulations~\cite{Bernhard:2019bmu}.
For the results with thick neutron skin contributions $\rnp=0.28$ fm, the predictions for $v_{3}\{2\}$ become better, but the $v_{2}\{2\}$ and $v_{2}\{4\}$ data are underestimated. Here, we did not change the previous parameters~\cite{Bernhard:2019bmu} obtained from Bayesian analysis to further improve the description of individual flow harmonics, but the results indicate that the predictions of flow observables from hydrodynamic simulations can be improved by considering the effect of neutron skin thickness.

\begin{figure}[htb]
	\begin{centering}
        \includegraphics[width=0.44\textwidth]{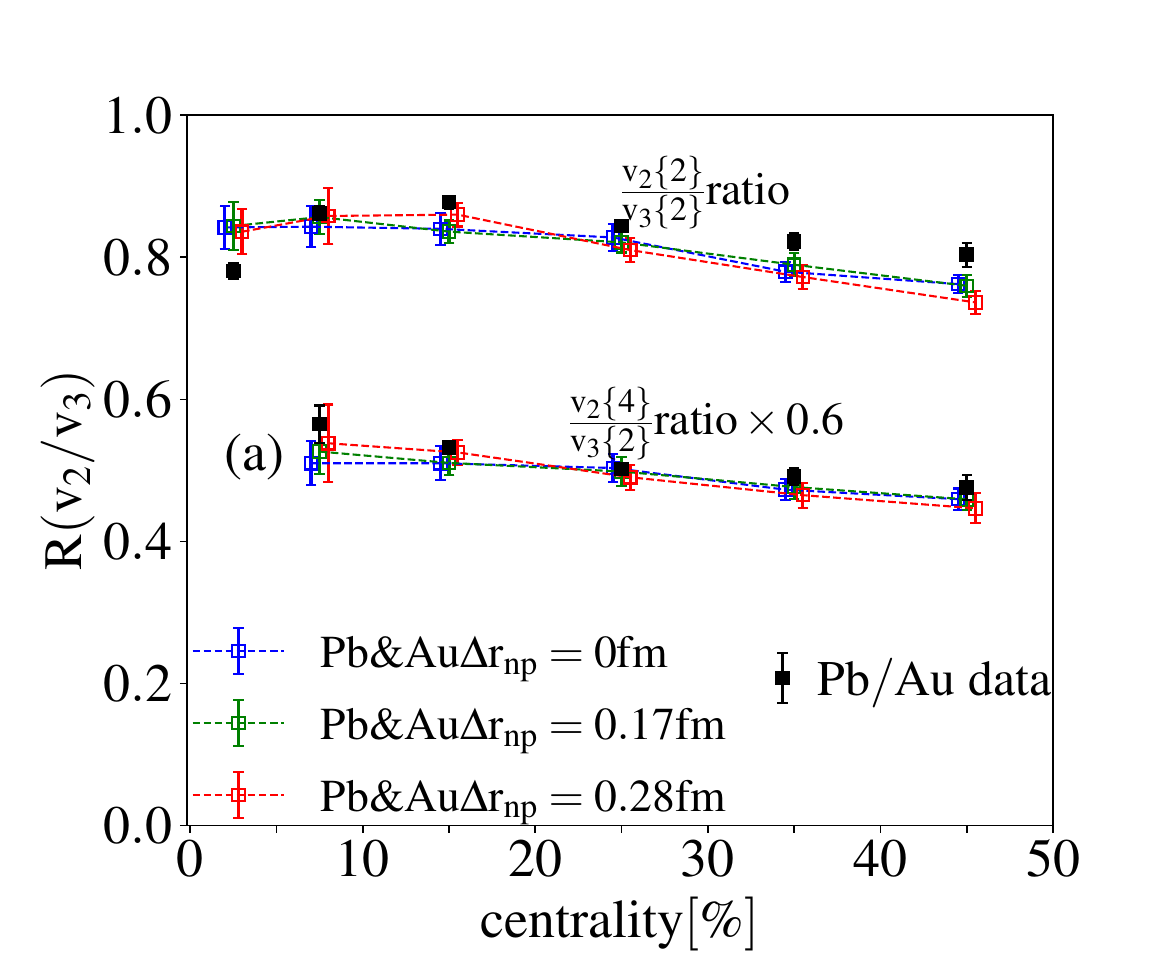}
	\end{centering}
	\caption{(Color online) The approximate scaling law in the isobaric ratio of ${v_2\{2\},v_2\{4\},v_3\{2\}},v_2\{2\}/v_3\{2\},v_2\{4\}/v_{3}\{2\}$ when \Pb{} and \Au{} have same neutron skin size.}
    \label{fig:scalinglaw}
\end{figure}

Figure~\ref{fig:v2v3} shows the ratios of $v_{2}\{2\}/v_{3}\{2\}$ and $v_{2}\{4\}/v_{3}\{2\}$ in both Au+Au collisions and Pb+Pb collisions.
The geometry and fluctuations of the diffuseness parameter $a$ are crucial for the simulation of the flow observables in relativistic heavy ion collisions~\cite{Xu:2021vpn,Wang:2023yis}.
In general, the differences in $a$ between charge and mass densities are mostly due to the halo-type neutron skin thickness in nuclear structure theory calculations.
For most of the central collisions, the ratios of $v_{2}\{4\}/v_{2}\{2\}$ are smaller in Au+Au collisions than in Pb+Pb collisions (not shown), indicating that the fluctuation contributions are larger in Au+Au collisions than in Pb+Pb collisions because the system size is smaller in Au+Au collisions with smaller mass number.
The differences between the predictions with different sizes of the neutron skin become even more obvious due to the opposite contributions at $v_{2}\{2\}$ ($v_{2}\{4\}$) and $v_{3}\{2\}$.
With the current parameter set, the ratio observables from most-central to semi-central collisions can be well explained by the hydrodynamic simulations with a neutron skin of $\rnp=0.17$ fm for both \Pb\ and \Au.

The individual flow observables are also sensitive to the bulk properties of the QGP medium. The ratios of flow observables in a single collision system discussed above can reduce the systematic uncertainties and the uncertainties from bulk evolution to some extent.
However, this cancellation in a single collision system is not good enough.
For example, we have checked that the observables discussed above are sensitive to the Gaussian smearing parameter $w$ in the initial \trento\ simulations.
To further reduce or eliminate these uncertainties, one can study the flow observables in two collision systems with similar system size and collision energy, as has been done for isobaric \RuRu\ and \ZrZr\ collisions.
%With similar collision energy, the semi-isobaric collision of Au+Au and U+U collisions can also be used to study the effect of nuclear deformations~\cite{Jia:2021wbq}.
Here we use Pb+Pb collisions and Au+Au collisions discussed above as semi-isobaric collisions, regardless of the differences in system size and collision energy.
Since the observables at top RHIC and LHC energies can be described by the hydrodynamical simulations with a fixed set of parameters~\cite{Schenke:2020mbo}, the uncertainties from the bulk properties would be small.
For example, we have checked that a $20\%$ difference in the free-streaming time $\tau_{f}$ between RHIC and LHC energies will not change our conclusion.
Of course, it is expected that the uncertainties can be further reduced when the data of the two systems at the same collision energy can be produced in the future.

Figure~\ref{fig:isobarratio}(a) shows the individual flow ratios, $R(v_{2}\{2\})$, $R(v_{2}\{4\})$ and $R(v_{3}\{2\})$, in the semi-isobaric collisions.
Here we have fixed the neutron skin for Au with $\rnp=0.17$ fm and changed the neutron skin for Pb.
A larger neutron skin in Pb gives larger negative contributions to $v_{2}\{2\}$ and $v_{2}\{4\}$, but a positive contribution to $v_{3}\{2\}$.
The results with different sizes of the neutron skin can be distinguished in the ratios of the individual flows, while the uncertainties are large for the model study with current statistics.
The double ratios of $R(v_{2}\{2\}/v_{3}\{2\})$ and $R(v_{2}\{4\}/v_{3}\{2\})$ can be used to enhance the significance of the differences.
The results shown in Fig.~\ref{fig:isobarratio}(b) indicate that the experimental data can be well described if Pb has the same size of neutron skin.
Here we have fixed the neutron skin for Au with $\rnp=0.17$ fm and changed the neutron skin of Pb.
One would expect the same conclusions if the neutron skin of Pb is fixed and the neutron skin of Au is tuned.

We have showed that the neutron skin thickness can be extracted from the flow ratio observables in relativistic semi-isobaric collisions when the neutron skin of the partner nuclei is known.
In the more general case, the neutron skin thicknesses of both partner nuclei are unknown.
The semi-isobaric collision can be used to study the relative differences between the two nuclei.
We find that if the Pb and Au nuclei have the same magnitude of neutron skin thickness, the conclusions from the double ratio flow observations do not change regardless of the exact value of the neutron skin chosen.
Such a scaling behavior is shown in Fig.~\ref{fig:scalinglaw}. All predictions follow the same trend when \Au\ and \Pb\ have the same size of neutron skin.
Based on this scaling law, the large difference in neutron skin thickness between \Pb\ and \Au\ measured by the PREX-II and STAR Collaborations is disfavored by measured flow observables and our calculations, indicating that semi-isobar collisions can provide some tight constraints on the neutron skin thickness and thus the nuclear symmetry energy at the relativistic collision energy.

\section{Conclusion}\label{sec:summary}
This paper focuses on studying the effect of the neutron skin from the flow observables measured in \PbPb\  and \AuAu\ collisions. Within the framework of iEBE-VISHNU hydrodynamic simulations,  the neutron skin thickness are introduced into the Woods-Saxon distribution of \Pb\ and \Au\ nuclei in the \trento\ initial condition. Using \dr(=$0.17$ fm)  extracted from $\rho^0\rightarrow\pi^+\pi^-$ photoproduction in ultra-peripheral Au+Au collisions, iEBE-VISHNU can roughly describe the flow data in \AuAu\ collisions at $\snn= 200$ GeV, together with the standard hydrodynamic parameters obtained from earlier Bayesian analysis and
an appropriate tuned normalization factor for initial entropy at top RHIC energy. Meanwhile, it demonstrates that larger neutron skin suppress the elliptic flow but increases the triangular flow
at both RHIC and LHC energies.  In order to partially remove the uncertainties from the bulk evolutions,
we also calculated the flow ratios $v_{2}\{2\}/v_{3}\{2\}$ and $v_{2}\{4\}/v_{3}\{2\}$ and found that
these two ratios in both Pb+Pb and Au+Au collisions can be well described by the hydrodynamics calculated with $\Delta r_{\rm np}=0.17$ fm for both \Pb\ and \Au\ nuclei.

To further investigate the effect of the neutron skin on the flow observables with less systematic uncertainty, we assumed the \PbPb\ collisions at LHC energy and the \AuAu\ collisions at top RHIC energy as semi-isobaric systems and calculated the flow ratio observables in such semi-isobaric collisions, despite the difference in collision energy and mass number. The associated flow ratio $R(v_{2}\{2\}/v_{3}\{2\})$ and $R(v_{2}\{4\}/v_{3}\{2\})$ between Au+Au and Pb+Pb collisions suggest that \Pb\ and \Au\ should have the same magnitude of neutron skin thickness. Thus, the large difference in neutron skin thickness between \Pb\ and \Au\ measured by the PREX-II and STAR collaborations is disfavored by the flow data and our calculations in Au+Au and Pb+Pb collisions. In short, similar to the exactly isobar collisions with \RuRu\ and \ZrZr\ at the same collision energy, the semi-isobar collisions proposed in this study can also give constraints on the neutron skin thickness and symmetry energy slope parameters in an unconventional way with a set of the existing flow data.

\section*{Acknowledgements}
This work is supported in part by the National Natural Science Foundation of China under Grant
Nos. 12247107, 12075007, HJX is supported by the National Natural Science Foundation of China under Grant Nos. 12275082, 12035006, 12075085, and 12147101.

\bibliography{ref}

\end{document}